\documentclass[usegraphicx]{mn2e}
\begin{document}

\title[Comment]{NGC 2419 does not challenge MOND, Part 2}

\author[R.H. Sanders] {R.H.~Sanders\\Kapteyn Astronomical Institute,
P.O.~Box 800,  9700 AV Groningen, The Netherlands}

 \date{received: ; accepted: }

\maketitle

\begin{abstract}
I argue that, despite repeated claims of Ibata et al., the
globular cluster NGC 2419 does not pose a problem for
modified Newtonian dynamics (MOND).  I present
a new polytropic model with a running polytropic
index.  This model provides an improved representation
of the radial
distribution of surface brightness while maintaining a
reasonable fit to the velocity dispersion profile.  Although 
it may be argued that the
differences with these observations remain large compared
to the reported random errors,  there are several
undetectable systematic effects
which render a formal likelihood analysis irrelevant. I
comment generally upon these effects and upon the intrinsic 
limitations of pressure supported objects as tests of
gravity. 
\end{abstract}

Ibata et al. (I1) have claimed that observations of
the distant globular cluster NGC 2419 constitute
a severe challenge for modfied Newtonian dyanmics (MOND),
that this object is, in fact, a ``crucible'' for theories
of gravity.  This claim was based, primarily, upon
Newtonian and MONDian Michie models in which the
phase space distribution is assumed to be of 
specific form leading to a density cutoff at a finite
radius.  In my initial response to their paper,
I argued that these models were too
restricted and presented high $n$ polytropic MOND
models (S11), in which the radial velocity dispersion 
decreases with radius; to the eye
in any case, such models appear to be quite consistent with
both the observed radial distribution of surface brightness
and the radial profile of velocity dispersion in this cluster. 
Further, I cautioned that it is questionable to rely on 
formal errors in likelihood analyses of observations which may well 
contain observational or intrinsic systematic effects.

Ibata et al. responded with their own analysis of polytropes
(I2) and assert that they had already, in effect, considered,
and ruled out, a range of non-isothermal MOND models which
fit the data better than the polytropic model that I
presented; that, in any case, Newtonian Michie models 
fit the observations significantly better than the MOND non-isothermal
models.  Again this conclusion is based upon the use of
random errors in a likelihood analysis.  It is primarily
this point that I wish to address here because it has
general relevance to the interpretation of 
astronomical observations.  But first, I present a new 
model in which the polytropic index runs with radius.

I2 point to the problem for polytropic MOND models as
a tension between the predicted 
surface density distribution
and the velocity dispersion radial profile.  This can be
summarized, in my own words, as follows:  to more precisely
fit the fall off of surface density within the very small
formal errors, in particular near
the outer cutoff, the polytropic
index must be rather large ($n > 15$); that is, the system
should be closer to isothermal.  But to fit the radial decrease
in the line-of-sight velocity dispersion, the index should be smaller
($n\approx 10$).  This result is fairly insensitive
to the anisotropy radius -- the distance beyond which radial
orbits begin to dominate.  

One obvious solution to this problem
is to allow the polytropic index to increase with radius
(there is nothing sacred about the rigid polytropic
pressure-density relation).  Therefore, I have considered several prescriptions
in which $n$ increases beyond 100 pc or so, for example,
$$n = n_0 \exp[r/r_p]. \eqno(1)$$  
Below I show the results for one such model.  Fig.\ 1
is the observed surface density of stars as a function
of radius (points) and that predicted by the
model (solid curve), and Fig.\ 2 is the same for the
line-of-sight velocity dispersion.  For this model
the central radial velocity dispersion is taken
to be 7.7 km/s, the central density is 38 M$_\odot$/pc$^2$,
the anisotropy radius is 11.5 pc, the central
polytropic index is 9.2 and the scale length for 
the growth of the polytropic index ($r_p$) is 850 pc.
Note that with a variable polytropic index, the
structure equation (Jeans equation) for the run
of density contains an
additional term reflecting the gradient in $n$.
The total mass of this model is $6.04 \times 10^5$
M$_\odot$ yielding a mass-to-light ratio of 1.4.
The value of MOND acceleration parameter 
is $a_0 = 10^{-8}$ cm/s$^2$ and the MOND interpolation
function was taken to be standard form as in S11.

Now, is this an acceptable match to the observations?
It is clearly an improvement over my initial polytropic
MOND model with a fixed $n$, particularly in
matching the surface density distribution while 
maintaining  a reasonable representation of
the velocity dispersion profile.  But Ibata et al. 
will probably
tell us, based upon a maximum likelihood analysis, 
that it is completely ruled out by the
observations or that it is 138 times less probable than
the best Newtonian Michie model.  Such claims,
of course, rest on the fact that the differences
between model and observations, while small to the eye,
are still larger than the reported 
measurement errors (comparable to the size of the
points in the surface density distribution).  But these
statements are misleading because they give an impression
of precision that is not, and {\it{cannot}} be, present
in astronomical observations of one such object.

\begin{figure}
\resizebox{\hsize}{!}{\includegraphics{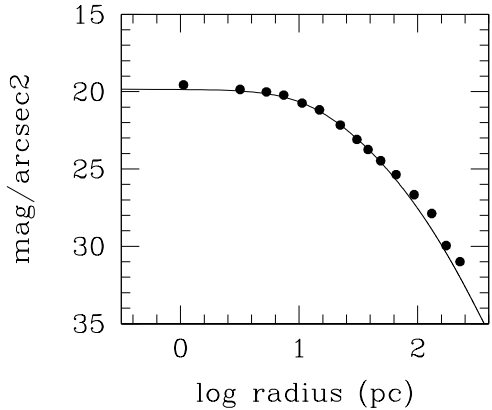}}
\caption{Surface brightness distribution in the
globular cluster NGC 2419 (points) compared to the
MOND non-isothermal (polytropic with running index), 
non-isotropic model described in the text.}
\label{}
\end{figure}

\begin{figure}
\resizebox{\hsize}{!}{\includegraphics{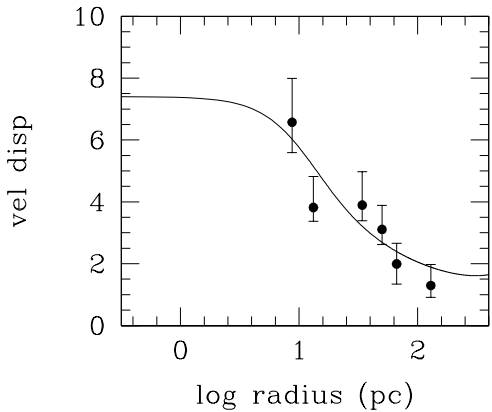}}
\caption{The observed radial dependence of
the line-of-sight velocity dispersion in NGC 2419 (points)
compared to that of the MOND model shown above.
}
\label{}
\end{figure}

I2 point out that in spherical symmetry,
given the anisotropy factor of a model, $\beta(r)$,
of a model, there is a unique relationship between the run of 
radial velocity and the density distribution.  This
is certainly true, but the observed quantities are not
the radial velocity or the density but the projected
versions of these quantities.  We know very well that 
small errors in the 
surface density can translate into significant
differences in the true density and, hence, in the
line-of-sight velocity dispersion (I2 admit that there
are problems with de-projection particularly in the
outer regions where the errors are larger). 

But let us assume that I1,2 perfectly understand
the systematic and random errors in their observations,
and that these are vanishingly small.
Then there remains a number of ``known unknowns" --
or more precisely ``known unknowables", for example,
the symmetry of the cluster.  I1,2 argue that
the cluster appears to be quite round and that this 
supports the assumption of spherical symmetry.  
But if the cluster is
oblate or prolate with the symmetry axis lying near
the line of sight, then the cluster would still appear
to be round without being spherically symmetric.
This would clearly alter the relation between
the run of line-of-sight velocity dispersion and
the projected density distribution, easily by more
than the random measurement errors.

Then there
is the unknown of rotation.  I1 consider an earlier
specific claim of rotation and dismiss this as an effect of
small number statistics.  But they cannot dismiss
the possibility of rotation in general. Rotation is most
likely to be present in the outer regions where
a systematic velocity of about 4 km/s would be
sufficient to reduce the model velocity dispersion at the
outermost measured point in Fig. 2 to the
observed value.  If the rotation axis were within
30 degrees of the line of sight its projection
would be less than 2.0 km/s;  given the 
the lower surface density of stars near the cutoff,
it is difficult to believe
that this would be detectable even in the very careful
observations of I1.

The point is that conceivable but undetectable
intrinsic systematic effects vitiate the value of
likelihood analyses (I haven't even
mentioned the ``unknown unknowns" -- effects 
that we have not thought of).
If so, it would not the first time
that sophisticated statistical analyses have been
applied to astronomical observations dominated
by systematic effects. 

There are fewer ambiguities with galaxy rotation curves.  This
is essentially because a two-dimensional object is being projected
onto a two dimensional sky.  There is one component of velocity
and the vector most often lies in one plane.  
The optical appearance of the disk combined
with a two dimensional radial velocity field can
generally unambiguously determine the projection, or
at least reveal when problems -- such as non-circular
disks, non-planar
motion, warping or non-circular motion -- are present.
As is well-known MOND works extremely well in predicting
the shapes of rotation curves using the observed 
distribution of baryonic matter (Sanders \& McGaugh 2002)
But even so, the rotation curves in a few percent of 
these well-specified systems are not in agreement
with those predicted
by MOND, nor would we expect them to be.  A perfect
theory would not precisely predict the rotation
curves of all disk galaxies because of the intrinsic
uncertainties.  

It is worse for spheroidal pressure-supported
objects because of uncertainties due to projection --
shape, rotation -- as well as the degree and radial dependence
of anisotropy of the velocity 
field and, in relaxed systems 
such as globular clusters, the possibility of mass 
segregation and varying 
mass-to-light ratio .  If Ibata et al. had several
such objects, all showing the same discrepancies,
their case would be stronger, but with only one
cluster, the small differences between model
predictions and observations are hardly definitive.  
I point out that 
opposite conclusion has been reached by Scarpa et al. (2011)
based upon observations of distant globular 
clusters which appear to be inconsistent with Newton.
The point is that strong statements on the 
nature of gravity cannot be made from observations
of a single such cluster given the uncertainties
intrinsic in astronomical observations of a 
three-dimensional object projected onto two dimensions.

I am grateful to Moti Milgrom for helpful comments on a draft
of this paper, and I thank an anonymous referee of my
previous paper on this subject for suggesting the idea
of a running polytropic index.


\begin{thebibliography}{}

\bibitem [Ibata et al. 2011a] {} Ibata R., Sollima A., Nipoti C.,
  Bellazzini M., Chapman S.C., Delessandro E., 2011, ApJ,
  738, 186 (I1).

\bibitem [Ibata et al. 2011b] {} Ibata R., Sollima A.,
  Nipoti C., Bellazzini M., Chapman S.C., Dalessandro E.,
  2011, ApJ (I2, in press).

\bibitem [Sanders 2011] {} Sanders, R.H. 2011, MNRAS,
  (in press, S11).

\bibitem [Sanders \& McGaugh 2002] {} Sanders R.H., McGaugh, S.S.,
  2002, Ann.Rev.Astron.Astrophys., 40, 263.

\bibitem [Scarpa et al. 2011] {} Scarpa R., Marconi G., Carraro G., 
  Falomo R., Villanova S., 2011, A\&A, 525, 148.

\end{thebibliography}
\end{document}